\documentclass[useAMS,usenatbib,usegraphicx]{mn2e} %
\usepackage{amsmath,amssymb}
\usepackage{psfrag}
\usepackage{rotating}
\usepackage{pstricks}
\usepackage{fixltx2e} % ensures the order of figures (one- and twocolumn) in the text
\newcommand{\be}{\begin{equation}}
\newcommand{\ee}{\end{equation}}
\newcommand{\bea}{\begin{eqnarray}}
\newcommand{\eea}{\end{eqnarray}}
\newcommand{\beao}{\begin{eqnarray*}}
\newcommand{\eeao}{\end{eqnarray*}}

\newcommand{\ds}{\displaystyle}
\newcommand{\nn}{\nonumber}

\title[Spectral energy distribution of super-Eddington flows]{Spectral energy distribution
of super-Eddington flows}
\author[D. Heinzeller et al.]{D. Heinzeller$^{1,2}$\thanks{E-mail:
dh@ita.uni-heidelberg.de}, S. Mineshige$^{2}$ and K. Ohsuga$^{3}$\\
$^{1}$Zentrum f\"{u}r Astronomie Heidelberg, Institut f\"{u}r Theoretische Astrophysik,
Albert-Ueberle-Stra{\ss}e 2, 69120 Heidelberg, Germany\\
$^{2}$Yukawa Institute for Theoretical Physics, Kyoto University, Kitashirakawa-Oiwakecho,
Sakyo-ku, Kyoto 606-8502, Japan\\
$^{3}$Department of Physics, Rikkyo University, 3-34-1 Nishi-Ikebukuro, Toshimaku, Tokyo 171-8501, Japan}
\begin{document}

\date{Accepted $\clubsuit$. Received $\clubsuit$; in original form $\clubsuit$}

\pagerange{\pageref{firstpage}--\pageref{lastpage}} \pubyear{$\clubsuit$}

\maketitle

\label{firstpage}

\begin{abstract}
Spectral properties of super-Eddington accretion flows are investigated by means of a
parallel line-of-sight calculation. The subjacent model, taken from two-dimensional
radiation hydrodynamic simulations by \citet{ohsuga_2005}, consists of a disc accretion
region and an extended atmosphere with high velocity outflows.
The non-gray radiative transfer equation is solved, including relativistic effects, by applying
the flux limited diffusion approximation.

The calculated spectrum is composed of a thermal, blackbody-like emission from the disc which
depends sensitively on the inclination angle, and of high energy X-ray and gamma-ray emission from the atmosphere.
We find mild beaming effects in the thermal radiation for small inclination angles.
If we compare the face-on case with the edge-on case, the average photon
energy is larger by a factor of $\sim 1.7$ due mainly to Doppler boosting, while
the photon number density is larger by a factor of $\sim 3.7$ due mainly to anisotropic
matter distribution around the central black hole. This gives an explanation for the
observed X-ray temperatures of ULXs which are too high to be
explained in the framework of intermediate-mass black holes.

While the main features of the thermal spectral component are
consistent with more detailed calculations of slim accretion discs, the atmosphere induces major
changes in the high-energy part, which cannot be reproduced by existing models.
We also conclude that, in order to interpret observational data properly, simple ap\-proa\-ches like
the Eddington-Barbier approximation cannot be applied.
\end{abstract}

\begin{keywords}
super-Eddington accretion -- spectral energy distribution -- beaming
\end{keywords}

\section{Introduction}\label{sec_introduction}
With the constant improvement of observational techniques, more and more detailed
information about accretion disc systems could be gained in the past decades.
In particular, a large number of ultraluminous X-ray sources (ULXs) have been discovered
since the end of the 1980's \citep{fabbiano_1989}. These new astrophysical objects imposed
a severe problem upon the existing general idea of accretion disc systems:
With a bolometric luminosity exceeding $10^{39}\,\textup{erg}\,\textup{s}^{-1}$ (derived from X-ray observations),
at least some of them show relatively low radiation temperatures ($\sim 0.1\,\textup{keV}$).
These systems have been suggested to be intermediate mass black hole (IMBH),
sub-Eddington accretion disc systems \citep{miller_2003,cropper_2004,roberts_2005}. However, from
stellar formation theory, it is difficult to explain the formation of IMBHs.
Therefore, many different formation theories have been presented in the last years.
A good overview can be found in \citet{marel_2004} and \citet{miller_2_2004}.
Controversially, from time-variability observations, it seems to be likely that these objects
are instead low-mass X-ray binary systems \citep{liu_2002}.
Radio observations show that the distribution of ULXs can as well be fitted
by stellar mass black holes with mildly relativistic jets \citep{koerding_2004}.

With increasing observational data from X-ray satellites like ASCA, Chandra and XMM-Newton,
large samples of ULX sources became available
(e.\,g. \citet{colbert_1999,makishima_2000,miller_2004,kubota_2006}).
They reveal that a distinct class of ULX sources exists, showing higher temperatures --
sometimes exceeding $1\,\textup{keV}$ --
than can be explained by IMBHs (e.\,g. \citet{okada_1998,mizuno_1999}).
Contrary, stellar mass Kerr black holes accreting above
their Eddington limit can account for these sources \citep{watarai_2001,ebisawa_2003}.
Alternatively, mild beaming could be important \citep{king_2001}, although the
shape of the surrounding nebula indicates that it is illuminated by a nearly isotropic
emission source \citep{wang_2002}. A controversial debate about the origin
of ULX sources is still ongoing.

Also in theory, a lot of investigations of accretion discs have been carried out
in the past decades, not at last due to enhanced computational facilities.
Systems hosting rotating black holes show higher temperatures
and enhanced luminosities due to a smaller inner boundary, compared to their non-rotating
counterparts \citep{gammie_1998,popham_1998}. To bypass the violation of the
Eddington limit for stellar mass black hole ULXs, various mechanisms have been proposed.
\citet{abramowicz_1988} extended the standard thin disc model \citep{shakura_1973} by
considering advective energy transport, leading to so called ``slim discs''.
These discs allow for higher mass accretion rates, compared to thin discs with the same luminosity.
The photon bubble instability mechanism permits higher accretion rates and high
luminosities while preserving the global stability of the disc \citep{begelman_2002}.
An important extension of the slim disc model is the full treatment of the photon
trapping effects (see \citet{katz_1977,begelman_1978} for the case of spherical accretion):
For sufficiently high accretion rates,
the gas inflow time becomes shorter than the photon diffusion time in the inner part
of the disc; high-energy photons otherwise emerging from the inner disc region are
partially swallowed by the black hole, leading to altered spectral energy distributions
and lower luminosities, compared to a standard slim disc \citep{ohsuga_2002,ohsuga_2003}.
The question about the validity of the Eddington limit in accretion discs
itself is widely discussed, yielding strong deviations for these disc models \citep{heinzeller_2006}.

Also the emerging disc spectra have been investigated widely in the past: Simple blackbody
or modified blackbody spectra for supercritical accretion discs have been calculated
\citep{szuszkiewicz_1996,wang_1999a,mineshige_2000,watarai_2000}. Furthermore, slim accretion disc spectra,
including self-irradiation and self-occultation for self-similar solutions
have been studied by \citet{fukue_2000}, while \citet{watarai_2005} investigated the implications
of geometrical effects and general relativistic effects on the disc spectra.
Further, \citet{kawaguchi_2003} considered Comptonization effects in spectral
calculations, finding significant spectral hardening occurring at large accretion rates.
In these approaches, however, anisotropy in radiation fields is not taken into account,
although we naively expect mild beaming effects, i.\,e., radiation is likely to escape
predominantly in the direction perpendicular to the disc plane. Moreover,
the influence of the environment of the disc, e.\,g. its atmosphere,
is not considered. It is well known that accretion discs own extended atmospheres, which indeed have
a strong influence on the emerging disc spectra. Photoionization of the accretion disc surface by
incident X-rays has been investigated by \citet{reynolds_1999}, while \citet{doerrer_1996}
calculated disc spectra for a thin $\alpha$-discs around a Kerr black hole, surrounded by a
hydrogen atmosphere.

In this study, we calculate the spectral energy distribution of supercritical accretion
flows based on the radiation hydrodynamic (RHD) simulations computed by \citet{ohsuga_2005}.
In Sect.~\ref{sec_model_setup}, we describe the methods of calculations used in this
investigation. Main aspects of the subjacent simulation data will also be briefly summarized
there. We then present our spectral calculations in Sect.~\ref{sec_results}.
Discussion will be given in Sect.~\ref{sec_discussion}, while Sect.~\ref{sec_conclusion}
is devoted to conclusions and an outlook.
\begin{figure}
\noindent\psfrag{O}{\begin{rotate}{-75}{{Obs.}}\end{rotate}}%
\noindent\psfrag{vacuum}{{{{vacuum}}}}%
\noindent\psfrag{comp}[c]{\rule{4mm}{0mm}{{{comp. area}}}}%
\noindent\psfrag{projected}[c]{\raisebox{2.5mm}{{{projected surface}}}\rule{4mm}{0mm}}%
\noindent\psfrag{T}{\rule{1mm}{0mm}\raisebox{2.5mm}{{{$\Theta$}}}}%
\noindent\psfrag{P}{\raisebox{0.2mm}{{{$\varphi$}}}}%
\noindent\psfrag{R}{\raisebox{0.2mm}{{{$r$}}}}%
\includegraphics[clip,width=0.95\columnwidth]{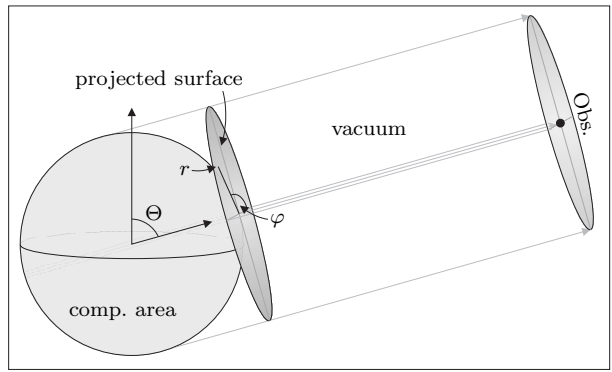}
\caption{Sketch of the line-of-sight calculation}\label{fig_sketch}
\end{figure}
\section{Model setup}\label{sec_model_setup}
\subsection{RHD simulations}\label{sec_rhd_model}
In this study, we account for both a sophisticated disc model and the disc's
atmosphere, computed in a self-consistent way within RHD simulations:
we apply our calculations to the 2D RHD simulation data from~\citet{ohsuga_2005}.
Starting with an empty disc and continuously injecting mass through the outer
disc boundary, the authors simulated the structure of a supercritical accretion flow,
until it reaches the quasi-steady state. The central object is given by a non-rotating stellar
mass black hole ($M=10 M_\odot$), generating a pseudo-Newtonian potential
\citep{paczynski_1980}. The viscosity is given by the classical $\alpha$ prescription.
The mass input rate at the outer boundary ($500 R_G$) strongly exceeds the
Eddington limit, $\dot{M}_\textup{ext} = 1000 \dot{M}_E$, where $\dot{M}_E = L_E/c^2$.
The authors considered energy transport through radiation and advection
and included relativistic effects in the radiation part. Note that
photon trapping effects were automatically incorporated in the simulations.
A gray computation of the radiative transfer in the flux limited
approximation~\citep{levermore_1981} was used.

They found that the supercritical flow is composed of two parts:
the disc region and the outflow regions above and below the disc.
Within the disc region, the circular motion as well as the patchy density
structure are observed. The mass accretion rate decreases inwards
(i.\,e. as matter accretes), roughly in proportion to the radius,
 and the remaining part of the disc material leaves the disc to form an outflow.
In particular, only $10\%$ of the inflowing material finally reach
the inner boundary ($3 R_G$), while the remaining $90\%$
get stuck in the dense, disc-like structure around the midplane
or transform into moderately high-velocity outflows with wide opening angles.
The outflows are accelerated up to $\sim 0.1 c$ via strong radiation pressure force.

From the simulation data, the gas density $\rho$, its temperature $T_\textup{gas}$
and its velocity $v$ are taken as input parameters, as well as the radiation energy density $E$.
The methods of calculating other quantities, such as the radiation
temperature $T_\textup{rad}$, the source function $S_\nu$ and the
radiation pressure tensor $P_\nu$, will be given in the following subsections.
\subsection{Equation of radiative transfer}\label{sec_rad_transfer}
Under the assumption of an observer being located at infinite distance from the object,
we calculate the emerging flux/luminosity as a function of the observers inclination angle $\Theta$
and azimuthal angle $\Phi$. Here, $\Theta$ and $\Phi$ refer to the spherical coordinate system which
describes the computational box.\footnote{Note that the equatorial plane of the
computational box is defined by the injection point of gas and its angular momentum vector.}

More precisely, we adopt a parallel line-of-sight calculation on a two-dimensional grid on
the projected surface, seen by the observer (see Fig.~\ref{fig_sketch} for a better
understanding). We start the line-of-sight calculation at a
sufficiently high optical depth $\tau_{\nu,\textup{start}}$
from the projected surface with initial intensity $I_\nu = 0$ and with fixed direction
cosine vector $\vec{l} = \vec{l}(\Theta, \Phi)$.

We focus on solving the radiative transfer equation numerically by a
Runge-Kutta algorithm, considering the relativistic
corrections due to the high velocities of the gas. General relativistic effects
such as gravitational lensing and gravitational redshift are not taken into account.
We have to distinguish between two different coordinate systems: We describe the system in which the
observer and the computational area are in rest with
$I_\nu$, $l$, etc., while the same quantities are tagged with $0$ in the
frame comoving with the local gas velocity $\vec{v}$.
Since the gas velocity strongly varies, the comoving frame
depends on the position in the simulation box.

In this framework, the relativistic equation of radiative transfer is given by
\bea
(\vec{l} \cdot \vec{\nabla}) I_\nu &=& \Bigl(\frac{\nu}{\nu_0}\Bigr)^2 \cdot
\Biggr\{\kappa^{\textup{abs}}_{\nu_0} S_{\nu_0} - \chi_{\nu_0}\,I_{\nu_0}\nn\\
&&\qquad+ \frac{3}{4} \kappa^{\textup{sca}}_{\nu_0} \frac{c}{4\pi} \left(E_{\nu_0} + l_{0i} l_{0j}
P_{\nu_0}^{ij}\right)\Biggl\}\label{eqn_rad_transfer_exact}\,,
\eea
while, in the non-relativistic case, it reduces to
\bea
(\vec{l} \cdot \vec{\nabla}) I_\nu &=& \Biggr\{\kappa^{\textup{abs}}_{\nu} S_{\nu} - \chi_{\nu}\,I_{\nu}\nn\\
&&\qquad+ \frac{3}{4} \kappa^{\textup{sca}}_{\nu} \frac{c}{4\pi} \left(E_{\nu} + l_{i} l_{j}
P_{\nu}^{ij}\right)\Biggl\}\label{eqn_rad_transfer_nonrel}\,.
\eea
It is important to note that all quantities on the right hand side of~\eqref{eqn_rad_transfer_exact}
are evaluated in the comoving frame, while those on the left hand side are given in the rest frame.
In the equation of radiative transfer, $S_\nu$ denotes the source function for matter,
$E_\nu$ the radiation energy density and $\mathbf{P}_\nu$ the radiation pressure tensor.
Additionally, $\chi_{\nu} = \kappa^{\textup{abs}}_{\nu} + \kappa^{\textup{sca}}_{\nu}$.

In this first approach, we restrict ourself to frequency-dependent absorption coefficients
for free-free absorption processes and totally neglect bound-free absorption processes,
$\ds \kappa^{\textup{abs}}_\nu = \kappa_{\nu}^\textup{ff}$. This holds as a good
approximation, since the gas temperature is mostly above $10^5\,\textup{K}$, and,
hence, hydrogen is fully ionized. For simplicity, we do not consider for metal opacities.
We adopt the formula given in \citet{rybicki_1979},
\be
\kappa_{\nu}^\textup{ff} = 3.7 \cdot 10^{8} T^{-1/2}
    \Bigl(\frac{\rho}{m_p}\Bigr)^2
    \nu^{-3} \left(1-e^{-h\nu/kT}\right)\,\textup{cm}^{-1}\,,
\ee
where we assume the mass density of ions and electrons to be
equal, $\ds \rho = \rho_i = \rho_e$. For the scattering processes, we only
consider electron scattering, given by

\be
\kappa^{\textup{sca}} = \sigma_T \Bigl(\frac{\rho}{m_p}\Bigr)\,\textup{cm}^{-1}\,.
\ee

\medskip\noindent
The relativistic transformation rules are
\bea
\nu_0&=&\nu \Gamma \Bigl(1-\frac{\vec{v} \cdot \vec{l}}{c}\Bigr)\label{eqn_transf_nu}\\
\vec{l}_0&=&\frac{\nu}{\nu_0} \left[\vec{l} + \left(c \frac{\Gamma-1}{{\vec{v}}^2} {\vec{v}}\cdot
            \vec{l}-\Gamma\right)\frac{\vec{v}}{c}\right]\label{eqn_transf_l}\\
I_{\nu_0}&=&\Bigl(\frac{\nu_0}{\nu}\Bigr)^3 I_\nu\label{eqn_transf_I}
\eea
with $\Gamma$ being the Lorentz factor.
\subsection{Frequency-dependent radiation quantities}\label{sec_further_requirements}
Special attention is needed for deriving the quantities $S_\nu$ and $E_\nu$ (and therefore, by
applying the flux limited diffusion approximation -- see Sect.~\ref{sec_fld} -- also $\mathbf{P}_\nu$).
As the radiative transfer in the 2D RHD simulation
is calculated in a {gray} approximation, the simulation data provides only frequency-integrated
values for the radiation energy density. The matter distribution is described by
the gas density and gas temperature. We assume local thermal equilibrium
for the matter distribution and for the radiation field separately:
\bea
S_{\nu_0}&=&B_{\nu_0}(T_\textup{gas})\ =\ \frac{2 h \nu_0^3}{c^2} \cdot
    \frac{1}{\exp\bigl(\frac{h\nu_0}{k_B T_\textup{gas}}\bigr) - 1}\label{eqn_Smatter}\\
T_\textup{rad}&=& \Bigl(\frac{E}{a}\Bigr)^{1/4},\quad a = \textup{ radiation constant}\label{eqn_Trad}\\
E_{\nu_0}&=&\frac{4\pi}{c}\,B_{\nu_0}(T_\textup{rad})\phantom{\int_0^1}\label{eqn_Erad}
\eea
While~\eqref{eqn_Smatter} generally holds as a good approximation, \eqref{eqn_Trad} and \eqref{eqn_Erad} have
to be treated carefully: In a scattering dominated domain as it is the case in the underlying
simulation data here, photons undergo multiple scattering and therefore
expand in space -- accordingly, the radiation field is diluted and the photon number decreases.
This implies that the average photon temperature will be underestimated by~\eqref{eqn_Trad}
and \eqref{eqn_Erad}, which will be discussed later (Sect.~\ref{sec_discussion}).
Therefore, we henceforth focus our discussion on the relative changes of the flux, photon energy and
photon number due to the variations of the inclination angle $\Theta$.
\subsection{Flux limited diffusion approximation}\label{sec_fld}
To calculate the radiation pressure tensor, we apply the frequency-dependent flux limited
diffusion (FLD) approximation \citep{levermore_1981} to the quantities in the comoving
frame. In this context, the radiation pressure tensor can be expressed by
\be
\mathbf{P}_{\nu, 0} = \mathbf{f}_\nu \,E_{\nu, 0}\,,\label{eqn_P_FLD}
\ee
where $\mathbf{f}_\nu$ is called the Eddington-tensor. Its components are given by
\be
f_\nu^{ij} = \frac{1}{2} (1-f_\nu) \delta^{ij} + \frac{1}{2} (3f_\nu-1)n_\nu^{i}n_\nu^{j}\,.\label{eqn_f_FLD}
\ee
Here, $n^{i}$ denotes the normalized energy density gradient,
\be
n_\nu^{i} = \frac{\bigl(\vec{\nabla} E_\nu\bigr)^{i}}{\bigl|\vec{\nabla} E_\nu\bigr|}\,.\label{eqn_norm_grad_E}
\ee
Following \cite{kley_1989}, $n_\nu^{i}$ and subsequent quantities can be
expressed as functions of the energy density in the inertial frame.

To close the resulting equations, the Eddington factor $f_\nu$ has to be determined.
From the momentum equations, the relation between $f_\nu$ and $\lambda_\nu$ is given by
\be
f_\nu = \lambda_\nu + \lambda_\nu^2 \mathcal{R}_\nu^2\,,\qquad \mathcal{R}_\nu = \frac{\bigl|\vec{\nabla} E_\nu\bigr|}{\chi_\nu E_\nu}\,.
\label{eqn_edd_factor_flux_limiter}
\ee
The flux limiter $\lambda_\nu$ itself can not be determined from the equations of radiative transfer,
but has to be defined manually. In order to do so, two conditions have to be fulfilled. In the
case of $\chi_\nu \to \infty$, the equations have to reduce to the classical diffusion limit,
i.\,e. $\lambda_\nu \to \frac{1}{3}$. In the case of $\chi_\nu \to 0$, the flux limiter must tend towards
$1/\mathcal{R}_\nu$ in order to ensure $|\vec{F}_{\nu, 0}| \leq c E_{\nu, 0}$.

Naturally, there exist multiple possibilities to describe the flux limiter $\lambda_\nu$. We
adopt the common formulation from~\citet{levermore_1981}:
\be
\lambda_\nu(\mathcal{R}_\nu)=\frac{2+\mathcal{R}_\nu}{6+3\mathcal{R}_\nu+\mathcal{R}_\nu^2}\label{eqn_flux_limiter}
\ee
\subsection{Numerics}
In the present investigation we choose a \emph{single} snapshot of the RHD simulation
data after the simulation settled down in a quasi-steady structure. In this stadium,
the structure does not change anymore in time in a significant way, giving rise to consider
our results as characteristic properties of such a system.\footnote{For details, we refer the
reader to \citet{ohsuga_2005}.}

The calculation of the disc spectra is performed as presented above with and without
relativistic corrections. As the simulation data is symmetric with respect to
the azimuthal angle $\Phi$, we compute the spectra only in dependency of the inclination
angle $\Theta$. We investigate the results of the computation for different starting points (i.\,e.,
optical depths $\tau_{\nu,\textup{start}}$) for the line-of-sight calculation. For
$\tau_{\nu,\textup{start}} \approx 8$, the results begin to saturate, leading to changes
below $1\%$ when starting at higher optical depths.
We used $\tau_{\nu,\textup{start}}=10$ throughout all cases and validated the results
with several integrations from $\tau_{\nu,\textup{start}}=10$, yielding differences below $10^{-3}$.

The reason why such a low $\tau_{\nu,\textup{start}}$ reveals the same results than
higher optical depths can be understood from the two extreme cases:
\begin{enumerate}
\item the gas is dense and cool with $T_\textup{gas} \approx T_\textup{rad}$ and
      small contributions of the gas to the total emissivity
\item it is diluted such that $\kappa^{\textup{abs}}_{\nu} \ll \kappa^{\textup{sca}}_{\nu}$
      and the total emission along the line of sight is completely dominated by the radiation field
\end{enumerate}
For all calculations it turns out that increasing gas temperatures go hand in hand with
dropping gas densities so that the opacity and total emission are governed by the radiation field.

We divide the projected surface seen by the observer in a polar grid with coordinates
$(r,\varphi)$, see Fig.~\ref{fig_sketch}. Both for the radial and the polar coordinate, we adopt a linear grid with
$N_r = 100$ and $N_\varphi=200$ grid points. The discretization in frequency is taken to
be logarithmic with $N_\nu=200$ frequency values between $10^{14}\,\textup{Hz}$ ($0.5\,\textup{eV}$) and
$10^{22}\,\textup{Hz}$ ($50\,\textup{MeV}$).
\begin{figure*}
\noindent\psfrag{nuLnu}{\begin{rotate}{-90}%
\raisebox{2mm}{$\nu L_\nu$}\end{rotate}}%
\noindent\psfrag{eg}[c]{$E_\gamma [\textup{eV}]$}%
\includegraphics[clip,width=0.8\textwidth]{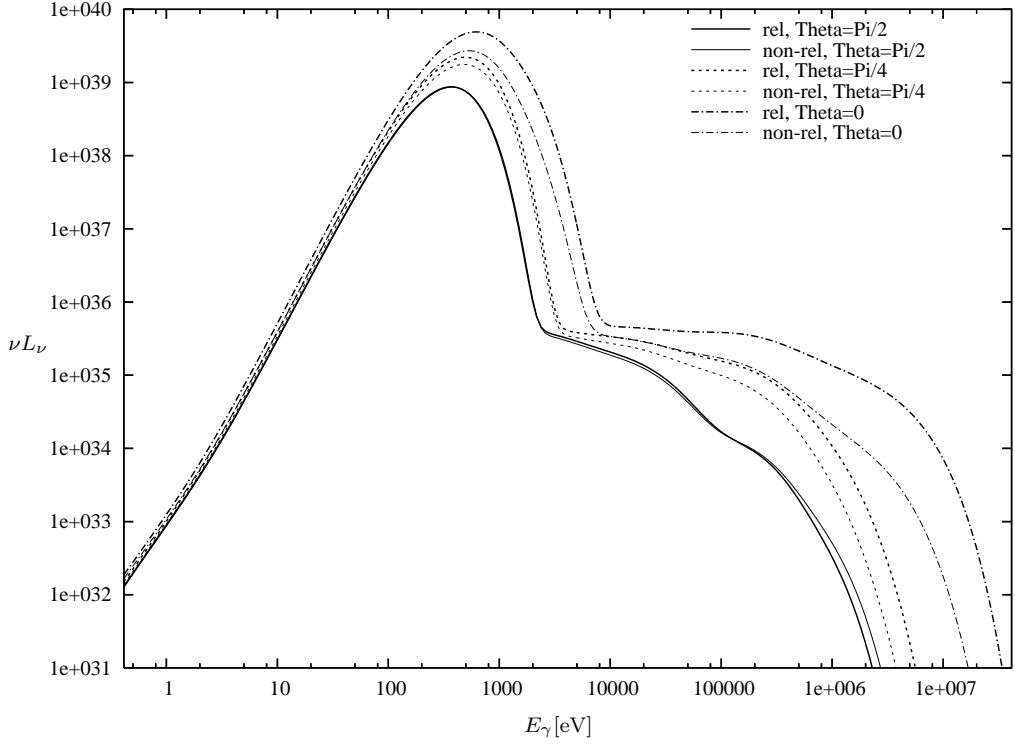}
\caption{Disc spectra $\nu L_\nu$ for inclination angles $\Theta = 0, \pi/4, \pi/2$ with and without
relativistic corrections}\label{fig_spectrum}
\end{figure*}
\section{Results}\label{sec_results}
\subsection{Overall spectral properties}
Figure~\ref{fig_spectrum} shows the resulting spectrum $\nu L_\nu$ for
inclination angles $\Theta = 0, \pi/4$ and $\pi/2$ with and without relativistic corrections.
The luminosity is given by
\be
L_\nu (\Theta) = 4 \pi \int_A I_\nu(\Theta, r, \varphi)\,dA\,,
\ee
where $A$ denotes the projected surface of the computational area, as it is seen by the observer.
For low frequencies, the spectra only weakly depend on the viewing angle.
Also, relativistic corrections are unimportant for energies $\lessapprox 400\,\textup{eV}$
($\nu \lessapprox 10^{17}\,\textup{Hz}$).
Contrary, for higher energies, the dependency on the viewing angle becomes stronger.
For high inclinations, i.\,e. for an edge-on view on the system, relativistic corrections
still remain unimportant, while they become drastically visible for low inclinations, i.\,e.
for a nearly face-on view on the disc. For both the relativistic and the non-relativistic
case, an enhancement of the peak frequency and luminosity is observable for small inclination angles;
although, this boost is much stronger when considering relativistic corrections.
Furthermore, instead of a rapid drop of $\nu L_\nu$ for high energies, a slower decline up to a plateau-like
structure can be seen in all cases.

\begin{figure*}
\noindent\psfrag{nuLnu}{\begin{rotate}{-90}%
\raisebox{2mm}{$\nu L_\nu$}\end{rotate}}%
\noindent\psfrag{eg}[c]{$E_\gamma [\textup{eV}]$}%
\includegraphics[clip,width=0.8\textwidth]{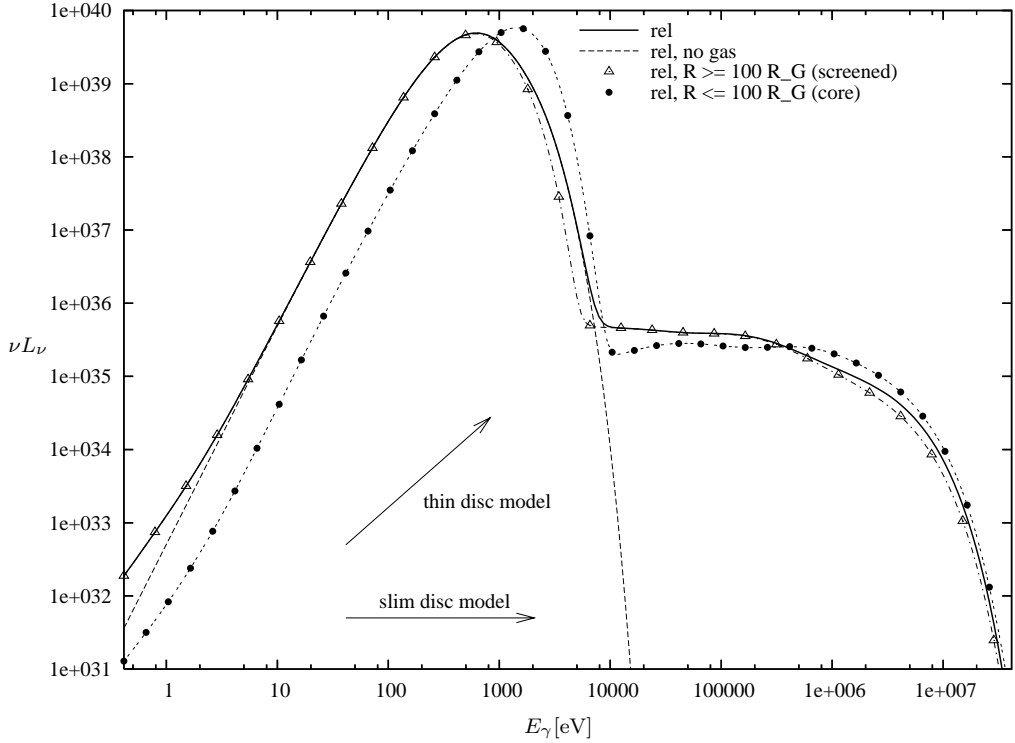}
\caption{Disc spectra $\nu L_\nu$ for $\Theta = 0$ with relativistic corrections. Beneath
the normal spectrum (as in Fig.~\ref{fig_spectrum}), we plot the spectra without gas contributions,
for the core region only ($R \leq 100 R_G$) and for a screened inner region ($R \geq 100 R_G$).
Additionally, theoretical spectral shapes for thin and slim accretion discs are sketched}\label{fig_realspectrum}
\end{figure*}
Figure~\ref{fig_realspectrum} illustrates again the spectrum for $\Theta = 0$. To show that
the observed plateau in Figs.~\ref{fig_spectrum} and~\ref{fig_realspectrum}
is a result of the thermal emission $\kappa^{\textup{abs}} S_\nu$ from the hot gas in the photosphere,
we calculate the spectrum without gas contribution, i.\,e. we set $S_{\nu,\textup{gas}}=0$ everywhere.
As it can be seen from the figure, the high-energy plateau disappears completely when neglecting
the contribution from the hot gas, confirming our hypothesis. We want to remark that
\begin{enumerate}
\item this structure may by altered significantly if Compton scattering is taken into account, since
this provides an effective cooling mechanism for the gas. However, the inclusion of Compton scattering
and the calculation of the decrease in $T_\textup{rad}$ is beyond the scope of this work.
\item it is unlikely to observe this plateau since the overall emission in this energy range is considerably low
and the spectrum is dominated by the peak emission.
\end{enumerate}
To illustrate the influence of the disc's environment
in the face-on case in Fig.~\ref{fig_realspectrum}, we calculate the spectrum for a ``screened'' central region (for $R \leq 100 R_G$, we set all
physical quantities to zero) and for the core region only (for $R \geq 100 R_G$, all
physical quantities are set to zero). Here, $R$ denotes the radial coordinate in the spherical
coordinate system describing the computational box. The former case corresponds to a system
where the inner $100 R_G$ are entirely evacuated and so emission, absorption and scattering
processes only exist outside the core region. Contrarily, the latter case means that no emission,
no absorption and no scattering takes place for $R \geq 100 R_G$, yielding the unaltered
emission from the core region only. Obviously, the environment of the disc
has a rather strong influence on the emerging spectrum.

Furthermore, the theoretical spectral shapes for a standard thin $\alpha$-disc \citep{shakura_1973}
and for a standard slim disc \citep{abramowicz_1988} are indicated in Fig.~\ref{fig_realspectrum}, each time without
consideration for self-irradiation, atmosphere, relativistic effects, etc.
By just taking into account the surface temperature $T_\textup{eff}$ for
these disc models and taking advantage of the face-on view (no self-occultation),
$\nu L_\nu \propto \nu^{4/3}$ for the thin disc case, and
$\nu L_\nu \propto \nu^{0}$ for the slim disc case (see, e.\,g., \citet[Sect.~3.2.5]{kato_1998}).
These shapes do not coincide with our results, although they should be valid at least for the
peak intensity region of the spectrum.
\begin{figure}
\noindent\psfrag{Ltot}{\begin{rotate}{-90}%
\raisebox{-1ex}{\small $L_\textup{tot}$}\end{rotate}}%
\noindent\psfrag{Theta}{\raisebox{-0.5ex}{\small $\Theta$}}%
\includegraphics[clip,width=0.95\columnwidth]{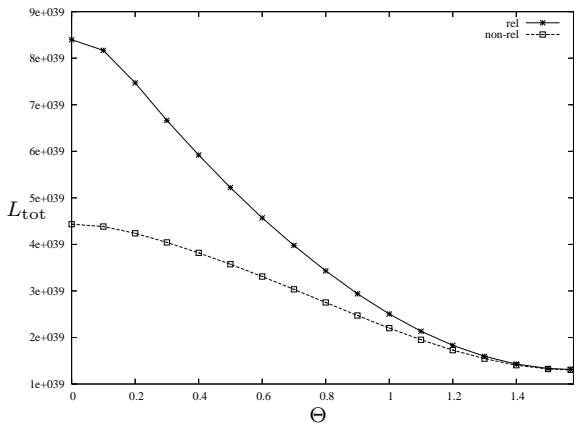}
\caption{Total luminosity $L_\textup{tot}$ as a function of the viewing angle
$\Theta$ for the relativistic and non-relativistic calculation}\label{fig_total_lum}
\end{figure}
\subsection{Angular dependence of the luminosity}
In Fig.~\ref{fig_total_lum}, we show the dependency of the total luminosity
\be
L_\textup{tot} (\Theta) = \int L_\nu(\Theta)\,d\nu
\ee
on the viewing angle for both the relativistic and the non-relativistic calculation.
The energy boost for small inclination angles appears in both cases, although
it is stronger for the relativistic calculation. Due to relativistic effects, the
gain in luminosity compared to the non-relativistic calculation $L_\textup{tot}^{\textup{rel}}/%
L_\textup{tot}^{\textup{non-rel}}$ varies between $1.0$ for $\Theta = \pi/2$ and
$1.9$ for $\Theta = 0$.

\begin{figure}
\noindent\psfrag{nph}{\begin{rotate}{-90}%
\raisebox{-7.5mm}{\small $n_\textup{tot}$}\end{rotate}}%
\noindent\psfrag{hnu}{\raisebox{3mm}{\begin{rotate}{-90}%
\raisebox{5mm}{\small $\langle h \nu \rangle$}\end{rotate}}}%
\noindent\psfrag{Theta}{\raisebox{-0.5ex}{\small $\Theta$}}%
\includegraphics[clip,width=0.95\columnwidth]{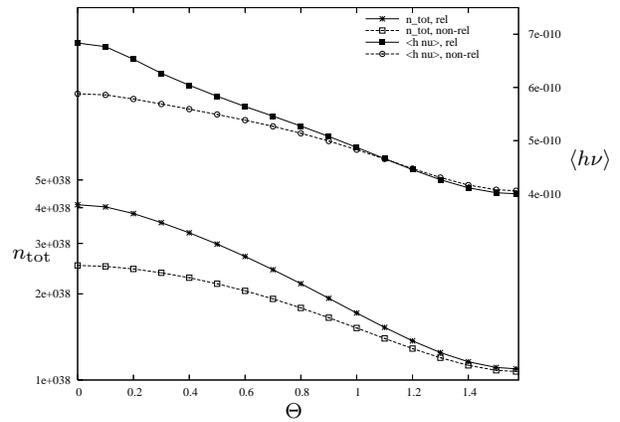}
\caption{Total photon number density $n_\textup{tot}$ and average photon energy $\langle h \nu \rangle$
in the relativistic and non-relativistic case as functions of the inclination angle $\Theta$
(see text for details)}\label{fig_photon_number}
\end{figure}
We thus conclude that $L_\textup{tot}$ is by a factor of $\sim 6.4$,
at most, enhanced for a face-on observer, compared
with an edge-on observer.  The increase in total luminosity may
be due either to an increase  in photon number or an increase
in average photon energy. Which one is more important?

To answer this question and to outline the relativistic
effects more explicitly, we plot in
Fig.~\ref{fig_photon_number} the total photon number density $n_\textup{tot}$
and the average photon energy $\langle h \nu \rangle$ as a function of the inclination
angle. From our SEDs, we calculate the photon number density by
\be
n_\textup{tot} (\Theta) = \int n_{\nu}(\Theta)\,d\nu = \int \frac{L_\nu(\Theta)}{h \nu c}\,d\nu
\ee
and from that, the average photon energy by
$\ds \langle h \nu \rangle (\Theta) = L_\textup{tot}(\Theta)/(c n_\textup{tot}(\Theta))$.
While relativistic effects become more or less unimportant
in the the edge-on case, they cause an additional
increase both in total number of photons originating from the system and in
average photon energy in the face-on case. Table~\ref{tab_summary_gains}
summarizes the gain in total luminosity, photon number and average photon energy for
the face-on case -- this means, the ratio for each of the quantities for $\Theta=0$,
compared to inclination angles between $0$ and $\pi/2$.
\begin{table}
\caption{Gain in total luminosity, photon number and average photon energy for
the face-on case (see text for details)}\label{tab_summary_gains}
\begin{tabular}{rcccccc} \hline
$\Theta$ & \multicolumn{2}{c}{$L_\textup{tot}$} & \multicolumn{2}{c}{$n_\textup{tot}$}
         & \multicolumn{2}{c}{$\langle h \nu \rangle$}\\
 & rel. & non-rel. & rel. & non-rel. & rel. & non-rel.\\ \hline
0       & $1.00$ & $1.00$ & $1.00$ & $1.00$ & $1.00$ & $1.00$\\
$\pi/6$ & $1.66$ & $1.26$ & $1.41$ & $1.17$ & $1.18$ & $1.08$\\
$\pi/4$ & $2.40$ & $1.59$ & $1.86$ & $1.39$ & $1.29$ & $1.14$\\
$\pi/3$ & $3.63$ & $2.14$ & $2.53$ & $1.72$ & $1.43$ & $1.24$\\
$\pi/2$ & $6.40$ & $3.40$ & $3.74$ & $2.35$ & $1.71$ & $1.45$\\ \hline
\end{tabular}
\end{table}

The results given in Table~\ref{tab_summary_gains} can be explained physically as follows:
Starting from the \emph{non-relativistic} calculation, we find that
\begin{itemize}
\item lower densities and therefore less effective absorption and scattering in the photosphere
allow a deeper look into the hotter region for the face-on case, compared to the edge-on case.
Hence, the average photon energy $\langle h \nu \rangle$ is increased by a factor of $1.45$.
\item photons can escape more easily through the diluted medium along the polar axis, while they
get stuck in the dense disc-like structure concentrated in the midplane. The outflow is therefore
collimated and the number of escaping photons raised by a factor of $2.35$.
\end{itemize}
At the same time, the (outflow) velocities of the gas close to the black hole ($R \lessapprox  100 R_G$)
and around the polar axis are higher, which becomes important for the \emph{relativistic} calculation.
\begin{itemize}
\item The frequency of the escaping photons is shifted from $\nu$ to $\nu_0 \geq \nu$ by the relativistic
Doppler effect, increasing the average photon energy additionally by a factor of $1.71/1.45=1.18$, when
comparing the face-on with the edge-on view.
\item Since the relativistic invariant is $I/\nu^3$, the emerging intensity in the relativistic calculation,
compared to the non-relativistic case is given by $I_0/I \sim (\nu_0/\nu)^3$. One factor of $\nu_0/\nu$
directly goes into $\langle h \nu \rangle$ via the relativistic Doppler effect, the remaining factor of
$(\nu_0/\nu)^2$ applies to the emerging photon number $n_\nu \sim I_\nu/(h\nu)$, raising it once more by
factor of $3.74/2.35=1.59$, when $\Theta$ decreases from $\pi/2$ to $0$.
\end{itemize}
All in all, an observer located at $\Theta = \pi/2$ only sees the emission from the outer part of
the optically thick disc-like structure, which itself screens the relativistic effects in the inner
region of the system. The radial velocities and also the azimuthal velocities are relatively low
($v_\varphi \approx 0.01 c$). For the mainly contributing part to the spectrum, the azimuthal
velocity is (almost) perpendicular to the line of sight, therefore the already weak relativistic effects are
not detectable for an edge-on observer. In the face-on case, the highly relativistic flow
($v_\textup{s} \lessapprox 0.3 c$) can by observed due to the optically thin atmosphere
above the disc. At the same time, the radial velocity is pointing into the direction of the observer,
leading to strong enhancements of the radiative flux at low inclinations.
\subsection{Blackbody fitting}
\begin{figure}
\psfrag{Trel}{\raisebox{2mm}{\begin{rotate}{-90}%
\raisebox{1mm}{\small $T_\textup{rel}$}\end{rotate}}}%
%\psfrag{epsfit}{\begin{rotate}{-90}%
%\raisebox{-10mm}{\small $\varepsilon$}\end{rotate}}%
\psfrag{Theta}{\raisebox{-0.5ex}{\small $\Theta$}}%
\includegraphics[clip,width=0.95\columnwidth]{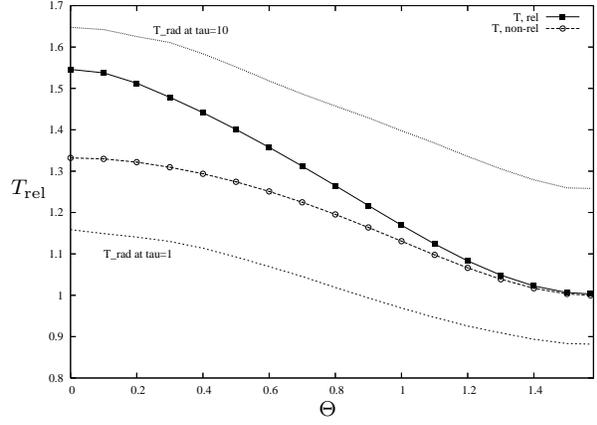}
%\caption{Fitted blackbody temperatures and spectral hardening factors
%for the relativistic and non-relativistic calculation as a function of
%the viewing angle $\Theta$. Additionally, mean temperatures at
%$\tau_\nu = 1$ and $\tau_\nu = 10$ are shown. All temperatures are colour-corrected, see text for details}\label{fig_eps_T_fit}
\caption{Fitted blackbody temperatures for the relativistic and
non-relativistic calculation as a function of the viewing angle
$\Theta$. Additionally, mean temperatures at $\tau_\nu = 1$ and
$\tau_\nu = 10$ are shown. All temperatures are scaled by $T_\textup{fit}$ at $\Theta=\pi/2$}\label{fig_eps_T_fit}
\end{figure}

When spectral data of binary black hole sources are
obtained, it is usual to fit them with blackbody (or disc
blackbody) spectra. We thus attempt a similar spectral
fitting to our theoretically calculated spectra:
We apply a non-linear least square fit to the emerging intensity $I_{\nu}$ by
a blackbody spectrum with temperature $T_\textup{fit}$, altered by a spectral
hardening factor $\varepsilon$ \citep{soria_2002}. The fitting function is then given by
\be
f = f(\nu, \varepsilon, T_\textup{fit}) = \varepsilon^{-4} \cdot \frac{2 h \nu^3}{c^2} \cdot
    \frac{1}{\exp\bigl(\frac{h\nu}{\varepsilon k_B T_\textup{fit}}\bigr)-1}\,.
\ee
Note that the factor $\varepsilon^{-4}$ is introduced to
ensure the same radiation energy loss:
\be
\pi \cdot \int f(\nu, \varepsilon, T_\textup{fit})\,d\nu =
\varepsilon^{-4} \sigma (\varepsilon T_\textup{fit})^4 = \sigma T_\textup{fit}^4
\ee
In order to account for stochastic fluctuations, we weigh the fitting
coefficients by their relative intensity. So, the weight-function is given by
\be
w(\nu) = \frac{I_\nu}{I_\textup{tot}},\qquad I_\textup{tot} = \int I_\nu\,d\nu\,.\label{eqn_weight_function}
\ee
Figure~\ref{fig_eps_T_fit} shows the results for the fitting temperatures
$T_\textup{fit}$ as a function of the inclination angle $\Theta$. Additionally,
we plot the surface-averaged radiation temperatures
\be
\bar{T}_\textup{rad} = \frac{1}{A} \int_{A} T_\textup{rad}\,dA
\ee
at optical depths $\tau_\nu = 1$ and $\tau_\nu = 10$. As mentioned earlier,
the temperature of the radiation field will be underestimated by
\eqref{eqn_Trad} and \eqref{eqn_Erad}. We therefore concentrate
on its relative changes for different inclinations and scale
all temperatures to the fitting temperature $T_\textup{fit}$ at
$\Theta=\pi/2$, where it is basically the same for the relativistic
and for the non-relativistic calculation.

If neglecting relativistic corrections, the fitted black\-body temperature is given roughly
by the radiation temperature at a \emph{constant} optical depth between $1$ and $10$.
The blackbody temperature rises by a factor $1.3$ when switching from an edge-on to a
face-on case. The spectrum is only weakly hardened compared to a Planck distribution at the
same temperature $T_\textup{fit}$: The spectral hardening factor $\varepsilon$ adopts an almost
constant value close to unity, $\varepsilon \approx 1.15$, for all inclinations.

When accounting for relativistic corrections, no surface
of {constant} optical depth can be defined any more: While the fitting temperatures
resemble those of the non-relativistic case for high inclinations, they differ significantly
for low inclinations, mirroring the above statement of stronger relativistic effects for
the face-on seen system. While the blackbody temperature rises by a factor of $1.6$, the
hardening factor $\varepsilon$ stays almost constant around $1.15$, like in the non-relativistic
calculation.

\begin{figure}
\psfrag{Lnu}{\begin{rotate}{-90}%
{\small $L_\nu$}\end{rotate}}%
\psfrag{eg}[c]{\raisebox{-0.5ex}{\small $E_\gamma [\textup{eV}]$}}%
\includegraphics[clip,width=0.95\columnwidth]{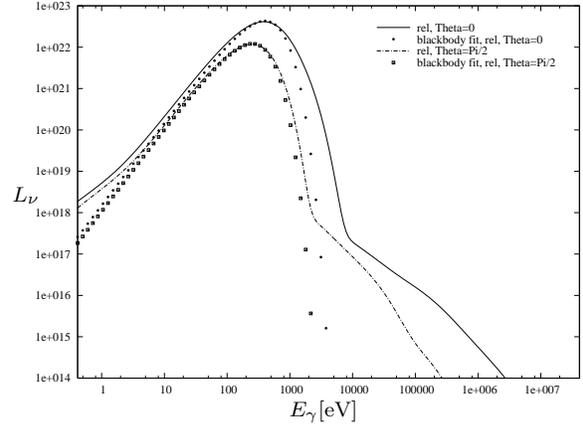}
\caption{Luminosity $L_\nu$ and the corresponding blackbody fits for
the face-on and edge-on case}\label{fig_comparison_result_fit}
\end{figure}

Finally, Fig.~\ref{fig_comparison_result_fit} shows the luminosity $L_\nu$
for the face-on and edge-on view and the corresponding blackbody fits in the relativistic case.
Due to the weighting function, the peak intensity region is fitted quite well, while there
are large deviations in the low-energy and high-energy part.
\section{Discussion}\label{sec_discussion}
The results presented in Sect.~\ref{sec_results} permit to draw several conclusions
about the observational appearance of supercritical accretion disc systems.

It is clearly not sufficient to consider only the disc
and neglect its surroundings like its hot photosphere, outflow regions, etc.
Their influence becomes most important in the high-energetic part of the spectrum
($h \nu \gtrapprox 4\,\textup{keV}$). We find a plateau-like structure, independent
of relativistic effects and of the viewing angle, which can be ascribed directly
to the high gas temperature in the corona. Therefore, neither the basic thin disc
spectrum, nor the basic slim disc spectrum fit to our results.

Our results also confirm that the Eddington-Barbier approximation,
a common simplification of radiative transfer calculations for stellar atmospheres,
cannot be applied in accretion discs:
In this approximation, one generally assumes that the emergent intensity along the
line of sight is equal to the source function at constant optical depth $\tau = 2/3$.
In our calculation, the main contribution to the emerging flux is produced at higher optical depths
$\tau_\textup{eff} > 2/3$; moreover, the exact value of $\tau_\textup{eff}$ depends
on the inclination angle.

We observe an enhanced luminosity for more and more face-on seen systems,
which is due to both enhanced average photon energy and total photon number.
Relativistic effects alter the total photon number much more significantly
(almost twice the non-relativistic treatment) than the average photon energy.
This can be identified as \emph{mild relativistic beaming}.

As mentioned above, due to \eqref{eqn_Trad} and \eqref{eqn_Erad}, the resulting
fit temperatures underestimate the real temperature of the radiation field.
Strictly speaking, the temperature of the radiation field should be determined
at $\tau^{\ast} \approx 1$ ($\tau^{\ast} \sim \sqrt{\tau^{\textup{sca}} \tau^{\textup{abs}}}$)
and not at $\tau \approx 1$ (in a scattering dominated domain, $\tau  \sim \tau^{\textup{sca}}$).
Hence, the radiation energy density $E_\nu$ will resemble more a shifted
black body distribution with a colour-corrected temperature $T_\textup{col}$,
\be
E_\nu \sim B_\nu(T_\textup{col}) \sqrt{{\kappa^{\textup{abs}}_\nu}/{\kappa^{\textup{sca}}}}\,,
\ee
rather then \eqref{eqn_Erad}, where $E_\nu \sim B_\nu(T_\textup{fit})$. From the requirement of energy conservation,
\be
E = \int B_\nu(T_\textup{fit})\,d\nu = % a T_\textup{fit}^4 \stackrel{!}{=}
\int B_\nu(T_\textup{col}) \sqrt{\frac{\kappa^{\textup{abs}}_\nu}{\kappa^{\textup{sca}}}} \,d\nu\,.
\label{eqn_energy_conservation}
\ee
To get a rough idea on how much the derived temperatures are underestimated,
we solve \eqref{eqn_energy_conservation} numerically for $T_\textup{col}$
in the main emanating region of radiation ($\tau \approx 10$). This yields
a correction factor $\xi = T_\textup{col}/T_\textup{fit} \approx 10$.
With
\be
T_\textup{fit} = \left[9.4 \cdot 10^5 \textup{K} \ldots 1.4 \cdot 10^6\textup{K}\right]
\ee
for $\Theta=\left[\pi/2 \ldots 0\right]$, this leads to colour-corrected temperatures in the range of
\be
T_\textup{col} = \left[9.4 \cdot 10^6 \textup{K} \ldots 1.4 \cdot 10^7\textup{K}\right]
\ee
These temperatures would be consistent with the observed
high temperatures of several ULX sources~\citep{makishima_2000}
that can not be explained in terms of intermediate mass black hole systems with
sub-Eddington accretion rates. However, our approach is certainly too simplified
to answer this ``too hot accretion disc'' puzzle in a satisfactory way.

In this work, spectral hardening turns out to be negligible. This may be due in parts
to the assumption of Thomson scattering: Comptonization effects are expected to harden the
spectrum significantly \citep*{czerny_1987,ross_1992,kawaguchi_2003}. Then, if only the
peak of the spectrum is observed, the absolute scale and therefore the
spectral hardening factor $\varepsilon$ remains
unknown and the observed temperature $T_\textup{obs} = \varepsilon T_\textup{col}$
overestimates the colour temperature $T_\textup{col}$.
{Moreover, bulk motion Compton scattering is known to alter photon energies
due to the angular redistribution of the scattered photons \citep{psaltis_1997}.
\citet*{socrates_2004} showed that turbulent Comptonization produces a significant
contribution to the far-UV and X-ray emission of black hole accretion discs.}

\begin{figure}
\noindent\psfrag{nuLnu}{\begin{rotate}{-90}%
\raisebox{1mm}{$\nu L_\nu$}\end{rotate}}%
\noindent\psfrag{eg}[c]{$E_\gamma [\textup{eV}]$}%
\includegraphics[clip,width=0.95\columnwidth]{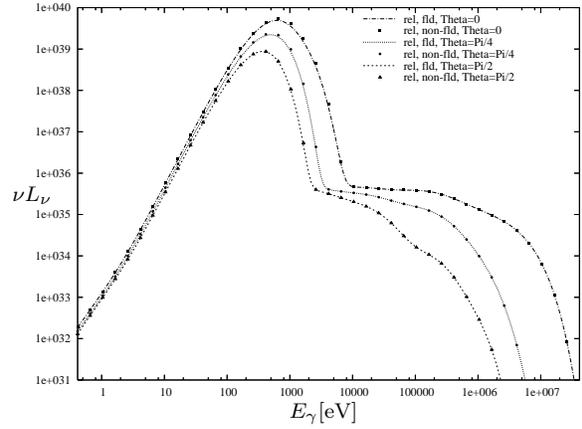}
\caption{Disc spectra $\nu L_\nu$ for $\Theta = 0, \pi/4, \pi/2$, computed under the
FLD approximation and complete isotropy}\label{fig_fld_contribution}
\end{figure}

A weak point in our investigation is the application of the flux limited
diffusion approximation instead of solving the full momentum equations: In this approximation,
several terms in the equation of radiative transfer~\eqref{eqn_rad_transfer_exact},
like $1/c^2\,(DF/Dt)$ with $F$ being the absolute value of the flux, are dropped. These terms are of the
order of $v/c$ and may contribute to the relativistic effects we find in our spectral
calculations. By calculating the emerging spectra under the classical diffusion limit
(i.\,e. complete isotropy, $\lambda=1/3$), we find only little influence of the FLD approximation at all
(see Fig.~\ref{fig_fld_contribution}). Thus, the inconsistencies invoked
by applying the flux limited diffusion approximation do not affect our results in a significant way.
\section{Conclusion and outlook}\label{sec_conclusion}
Our radiative transfer calculations, based on the 2D RHD simulation of highly accreting supercritical discs
including the photon trapping mechanism, show that the interpretation of observed disc spectra
is not a straightforward task. Especially, we find moderate beaming effects when the system
is viewed from nearly face-on; i.\,e., the average photon energy is larger by a factor of
$\sim 1.7$ in the face-on case than in the edge-on case due mainly to the Doppler boosting.
Likewise, the photon number density is larger by a factor of $\sim 3.7$ because of the anisotropic
matter distribution around the central black hole.
Interpreting observations thus has to be done in a more sophisticated way than one may expect
from basic disc models: It requires a careful treatment of the radiative transfer with
consideration for the discs surroundings.

We assume that both the gas and the radiation field separately
stay in local thermal equilibrium. Although the weak coupling of matter and
radiation ($\kappa_\textup{abs} \ll \kappa_\textup{sca}$) supports this assumption,
it remains questionable and also underestimates the temperature of the radiation field.
It is important to note that previous investigations by \citet{wang_1999b,fukue_2000,watarai_2005}
also rely on this approximation; nevertheless, their results differ in a significant way.
Solving the crux of assuming LTE for the gas and for the matter distribution at the present
stage is not possible, because it requires frequency-dependent RHD simulations.

In a next step, Compton scattering has to be included as well
as frequency-dependent absorption for both continuum (bound-free absorption may become
relevant in the low-energetic tail of the SED) and line processes:
From the observational side, emission lines, especially the K-shell transitions of iron,
are a prominent feature in accretion disc system and comprise many details about the
observed object (see~\citet{reynolds_1999,reynolds_2006} for example).
Beneath the effects on the high-energetic part of the spectrum mentioned before,
Compton scattering will provide an efficient cooling mechanism for the hot gas.

As pointed out by \citet{watarai_2005}, also general relativistic effects should be considered
in the vicinity of the black hole, which will primarily affect the spectra of face-on seen systems.
\section*{Acknowledgments}
The authors would like to thank Dr. Ken-ya Watarai, and Profs. Wolfgang J. Duschl, Jun Fukue and Shoji Kato
for useful comments and discussions. This work was supported in part by the Grants-in-Aid of the Ministry
of Education, Science, Culture, and Sport  (14079205, 16340057 S.M.), by the Grant-in-Aid for
the 21st Century COE ``Center for Diversity and Universality in Physics'' from the Ministry
of Education, Culture, Sports, Science and Technology (MEXT) of Japan, and by
the International Max Planck Research School for Astronomy and Cosmic Physics at
the University of Heidelberg.

\label{lastpage}

\newpage
\begin{thebibliography}{88.}
%\makeatletter
%\def\@biblabel#1{\hspace*{\labelsep}[#1]}
%\makeatother
\bibitem[\protect\citeauthoryear{Abramowicz et al.}{1988}]{abramowicz_1988} Abramowicz M.A., Czerny B.,
Lasota J.P., Szuzkiewicz E., 1988, ApJ, {332}, 646%--658

\bibitem[\protect\citeauthoryear{Begelman}{1978}]{begelman_1978} Begelman M.C.,
1978, MNRAS, {184}, 53%--67

\bibitem[\protect\citeauthoryear{Begelman}{2002}]{begelman_2002} Begelman M.C.,
2002, ApJ, 568, L97%--L100

\bibitem[\protect\citeauthoryear{Colbert \&\ Mushotzky}{1999}]{colbert_1999}
Colbert E.J.M., Mushotzky R.F., 1999, ApJ, {519}, 89%--107

\bibitem[\protect\citeauthoryear{Cropper et al.}{2004}]{cropper_2004}
Cropper M., Soria R., Mushotzky R.F., Wu K., Markwardt C.B., Pakull M., 2004, MNRAS, 349, 39%--51

\bibitem[\protect\citeauthoryear{Czerny \&\ Elvis}{1987}]{czerny_1987}
Czerny B., Elvis M., 1987, ApJ, 321, 305%--320

\bibitem[\protect\citeauthoryear{D\"{o}rrer et al.}{1996}]{doerrer_1996} D\"{o}rrer T., Riffert H.,
Staubert R., Ruder H., 1996, A\&A, {311}, 69%--78

\bibitem[\protect\citeauthoryear{Fabbiano}{1989}]{fabbiano_1989} Fabbiano G., 1989, ARA\&A, {27}, 87%--138

\bibitem[\protect\citeauthoryear{Fukue}{2000}]{fukue_2000} Fukue J., 2000, PASJ, {52}, 829%--840

\bibitem[\protect\citeauthoryear{Ebisawa et al.}{2003}]{ebisawa_2003} Ebisawa K.,
Zycki P., Kubota A., Mizuno T., Watarai K.-y., 2003, CHJAA, {3}, Suppl., 415%--424

\bibitem[\protect\citeauthoryear{Gammie \&\ Popham}{1998}]{gammie_1998} Gammie C.F., Popham R.,
%: Advection-dominated Accretion Flows in the Kerr Metric. I. Basic Equations,
ApJ {498} (1998), 313--326

\bibitem[\protect\citeauthoryear{Hayashi, Hoshi \&\ Sugimoto}{Hayashi et al.}{1962}]{hayashi_1962} Hayashi C.,
Hoshi R., Sugimoto D., 1962, Progr. Theoret. Phys. Supp., {22}, 1%--183

\bibitem[\protect\citeauthoryear{Heinzeller \&\ Duschl}{2006}]{heinzeller_2006} Heinzeller D.,
Duschl W.J., 2006, to appear

\bibitem[\protect\citeauthoryear{Kato, Fukue \&\ Mineshige}{Kato et al.}{1998}]{kato_1998}
Kato S., Fukue J., Mineshige S., 1998, Black-Hole Accretion Disks. Kyoto University Press, Kyoto

\bibitem[\protect\citeauthoryear{Katz}{1977}]{katz_1977}
Katz J.I., 1977, ApJ, 215, 265

\bibitem[\protect\citeauthoryear{Kawaguchi}{2003}]{kawaguchi_2003} Kawaguchi T., 2003, ApJ, {593}, 69%--84

\bibitem[\protect\citeauthoryear{King et al.}{2001}]{king_2001}
King A.R., Davies M.B., Ward M.J., Fabbiano G., Elvis M., 2001, ApJ, {552}, L109%--L112

\bibitem[\protect\citeauthoryear{Kley}{1989}]{kley_1989} Kley W., 1989, A\&A, {208}, 98%--110

\bibitem[\protect\citeauthoryear{K\"{o}rding, Colbert \&\ Falcke}{K\"{o}rding et al.}{2004}]{koerding_2004}
K\"{o}rding E., Colbert E., Falcke H., 2004, Progr. Theoret. Phys. Supp., {155}, 365%--366

\bibitem[\protect\citeauthoryear{Kubota \&\ Makishima}{2006}]{kubota_2006}
Kubota A., Makishima K., 2004, Advances in Space Research, Proc. of 35th COSPAR,
Paris, France, in press

\bibitem[\protect\citeauthoryear{Levermore \&\ Pomraning}{1981}]{levermore_1981} Levermore C.D., Pomraning G.C.,
1981, ApJ, {248}, 321%--334

\bibitem[\protect\citeauthoryear{Liu et al.}{2002}]{liu_2002} Liu J.-F., Bregman,
J.N., Irwin J., Seitzer P., 2002, ApJ, {581}, L93%--L96

\bibitem[\protect\citeauthoryear{Makishima et al.}{2000}]{makishima_2000}
Makishima K. et al., 2000,
% Kubota A., Mizuno T., Ohnishi T., Tashiro M., Aruga Y., Asai K., Dotani T.,
% Mitsuda K., Ueda Y., Uno S.'I., Yamaoka, K., Ebisawa, K., Kohmura, Y., Okada, K.,
ApJ, {535}, 632%--643

\bibitem[\protect\citeauthoryear{van der Marel}{2004}]{marel_2004} van der Marel R.P.,
2004, Intermediate-Mass Black Holes in the Universe: A Review of Formation
Theories and Observational Constraints. Carnegie Observatories Centennial Symposia,
Coevolution of Black Holes and Galaxies. Cambridge University Press, Cambridge

\bibitem[\protect\citeauthoryear{Miller et al.}{2003}]{miller_2003} Miller J.M., Fabbiano G., Miller,
M.C., Fabian A.C., 2003, ApJ, {585}, L37%–-L40

\bibitem[\protect\citeauthoryear{Miller et al.}{2004}]{miller_2004} Miller J.M., Zezas A., Fabbiano G.,
Schweizer F., 2004, ApJ, {609}, 728%--734

\bibitem[\protect\citeauthoryear{Miller \&\ Colbert}{2004}]{miller_2_2004} Miller M.C., Colbert E.J.M.,
2004, Int. J. Mod. Phys. D, {13}, 1%--64

\bibitem[\protect\citeauthoryear{Mineshige et al.}{2000}]{mineshige_2000}
Mineshige S., Kawaguchi T., Takeuchi M., Hayashida K., 2000, PASJ, {52}, 499%--508

\bibitem[\protect\citeauthoryear{Mizuno et al.}{1999}]{mizuno_1999}
Mizuno T., Ohnishi T., Kubota A., Makishima K., Tashiro M., 1999, PASJ, {51}, 663%--671

\bibitem[\protect\citeauthoryear{Ohsuga et al.}{2002}]{ohsuga_2002} Ohsuga K.,
Mineshige S., Mori M., Umemura M., 2002, ApJ, {574}, 315%--324

\bibitem[\protect\citeauthoryear{Ohsuga, Mineshige \&\ Watarai}{Ohsuga et al.}{2003}]{ohsuga_2003} Ohsuga K.,
Mineshige S., Mori M., Watarai K.-y., 2003,
ApJ, {596}, 429%--436

\bibitem[\protect\citeauthoryear{Ohsuga et al.}{2005}]{ohsuga_2005} Ohsuga K.,
Mori M., Nakamoto T., Mineshige S., 2005, ApJ, {628}, 368%--381

\bibitem[\protect\citeauthoryear{Okada et al.}{1998}]{okada_1998}
Okada K., Dotani T., Makishima K., Mitsuda K., Mihara T., 1998,
PASJ, {50}, 25%--30

\bibitem[\protect\citeauthoryear{Paczy\'{n}ski \&\ Wiita}{1980}]{paczynski_1980} Paczy{\'n}ski B.,
Wiita P.J., 1980, A\&A, {88}, 23%--31

\bibitem[\protect\citeauthoryear{Popham \&\ Gammie}{1998}]{popham_1998} Popham R., Gammie C.F.,
1998, ApJ, {504}, 419%--430

\bibitem[\protect\citeauthoryear{Psaltis \&\ Lamb}{1997}]{psaltis_1997} Psaltis D.,  Lamb F.K.,
1997, ApJ, {488}, 881

\bibitem[\protect\citeauthoryear{Reynolds}{2006}]{reynolds_2006}
Reynolds C.S., 2006, preprint (astro-ph/0605368)

\newpage

\bibitem[\protect\citeauthoryear{Reynolds et al.}{1999}]{reynolds_1999}
Reynolds C.S., Young A.J., Begelman M.C., Fabian A.C., 1999, ApJ, {514}, 164%--179

\bibitem[\protect\citeauthoryear{Roberts et al.}{2005}]{roberts_2005}
Roberts T.P., Warwick R.S., Ward M.J., Goad M.R., Jenkins L.P., 2005, MNRAS, 357, 1363%--1369

\bibitem[\protect\citeauthoryear{Ross, Fabian \&\ Mineshige}{Ross et al.}{1992}]{ross_1992}
Ross R.R., Fabian A.C., Mineshige S., 1992, MNRAS, 258, 189

\bibitem[\protect\citeauthoryear{Rybick \&\ Lightman}{1979}]{rybicki_1979} Rybicki G.B.,
Lightman A.P., 1979, Radiative processes in astrophysics. Wiley \&\ Sons, New York

\bibitem[\protect\citeauthoryear{Shakura \&\ Sunyaev}{1973}]{shakura_1973} Shakura N.I.,
Sunyaev R.A., 1973, A\&A, {24}, 337%--355

\bibitem[\protect\citeauthoryear{Socrates, Davis \&\ Blaes}{Socrates et al.}{2004}]{socrates_2004}
Socrates A., Davis S.W., Blaes O., 2004, ApJ, 601, 405

\bibitem[\protect\citeauthoryear{Soria \&\ Puchnarewicz}{2002}]{soria_2002} Soria R.,
Puchnarewicz E.M., 2002, MNRAS, {329}, 456%-–460

\newpage

\bibitem[\protect\citeauthoryear{Szuszkiewicz, Malkan \&\ Abramowicz}{Szuszkiewicz et al.}{1996}]{szuszkiewicz_1996}
Szuszkiewicz E., Malkan M.A., Abramowicz M.A., 1996, ApJ, {458}, 474%--490

\bibitem[\protect\citeauthoryear{Wang \&\ Zhou}{1999}]{wang_1999a} Wang J.-M., Zhou Y.-Y.,
1999, ApJ, {516}, 420%--424

\bibitem[\protect\citeauthoryear{Wang et al.}{1999}]{wang_1999b} Wang J.-M., Szuszkiewicz E.,
Lu F.-J., Zhou Y.-Y., 1999, ApJ, {522}, 839%--845

\bibitem[\protect\citeauthoryear{Wang}{2002}]{wang_2002} Wang Q.D.,
2002, MNRAS, {332}, 764%--768

\bibitem[\protect\citeauthoryear{Watarai et al.}{2000}]{watarai_2000}
Watarai K.-y., Fukue J., Takeuchi M., Mineshige S., 2000, PASJ, {52}, 133%--141

\bibitem[\protect\citeauthoryear{Watarai, Mizuno \&\ Mineshige}{Watarai et al.}{2001}]{watarai_2001}
Watarai K.-y., Mizuno T., Mineshige S., 2001, ApJ, {549}, L77%--L80

\bibitem[\protect\citeauthoryear{Watarai et al.}{2005}]{watarai_2005}
Watarai K.-y., Ohsuga K., Takahashi R., Fukue J., 2005, PASJ, 57, 513%--524
\end{thebibliography}
\end{document}